\documentclass[dvips]{article}

\usepackage{icrc2011}

\title{Extended X-ray jet and TeV emission in a low frequency peaked
   BL Lac object}

\newcommand{\etal}{\MakeLowercase{\textit{et al. }}} 
\shorttitle{author \etal paper short title}

\authors{S. Kaufmann$^{1}$, S. Wagner$^{1}$, O. Tibolla$^{2}$ }
\afiliations{$^1$Landessternwarte, Universit\"at Heidelberg\\ 
$^2$ITPA, Universit\"at W\"urzburg }
\email{s.kaufmann@lsw.uni-heidelberg.de}

\abstract{BL Lac objects are known to have very energetic jets pointing towards the observer
under small viewing angles. Many of these show high luminosity over the whole energy
range up to TeV, mostly classified as high-energy peaked BL Lac objects. Recently,
TeV gamma-ray emission was detected from a low-energy peaked BL Lac object.
Interestingly, this source has also a clear detection of an X-ray jet. We present a
detailed study of this X-ray jet and its connection to the radio jet as well as a
study of the underlying physical processes in the energetic jet, producing emission
from the radio to the TeV range. The orientation of the jet and the detected TeV
emission allow to constrain the parameters to describe the SSC emission and give the
opportunity to compare the influence of the viewing angle on the emission models with
other known TeV blazars.
}
\keywords{BL Lac object; Ap Lib; X-ray jet; multi-wavelength observations
}

\begin{document}
\maketitle

\section{Introduction}

BL Lac objects are known to have very energetic jets pointing under small viewing angle towards the observer. The spectral energy distribution (SED) of BL Lac objects shows two significant/prominent peaks. The lower frequency peak in the optical-UV-X-ray regime can be explained by synchrotron emission and the higher frequency peak at keV-GeV-TeV frequencies results from inverse Compton emission from a population of relativistic electrons which upscatter their self-produced synchrotron photons within the most common leptonic models, e.g. \cite{Maraschi1992}. Also alternative models describing instead hadronic interactions exist, e.g. \cite{Mannheim1993}.
Concerning the location of the emission peaks in the spectral energy distribution, the BL Lac objects are classified as high-energy, intermediate-energy and low-energy peaked. Most prominent for this classification is the slope in the X-ray spectrum. For the high-energy peaked BL Lac objects the X-ray spectrum describes part of the synchrotron emission, while for the intermediate-energy peaked it is part of the synchrotron and inverse Compton emission and represented by a flat spectrum in the SED (photon index around 2). For the low-energy peaked BL Lac objects (LBL), the X-ray emission is dominated by inverse Compton emission. Following the standard SSC models, it is therefore rather unusual to detect VHE (E$>$100GeV) emission from an LBL. 

\section{Ap Lib -- PKS 1514-241}

The low-energy peaked BL Lac object AP Librae is well known as one of the most active blazars in optical. It has a redshift of z=0.049 \cite{Woo2005}  and is located at $\alpha_{\rm{J2000}} = 15^{\rm{h}} 17^{\rm{m}} 41.8^{\rm{s}}$, $\delta_{\rm{J2000}} = -24^\circ 22' 19''$.
It has been classified as a BL Lac object by \cite{Strittmatter1972} and \cite{Bond1973}.

 \begin{figure*}[!t]
   \centerline{\includegraphics[width=0.48\textwidth]{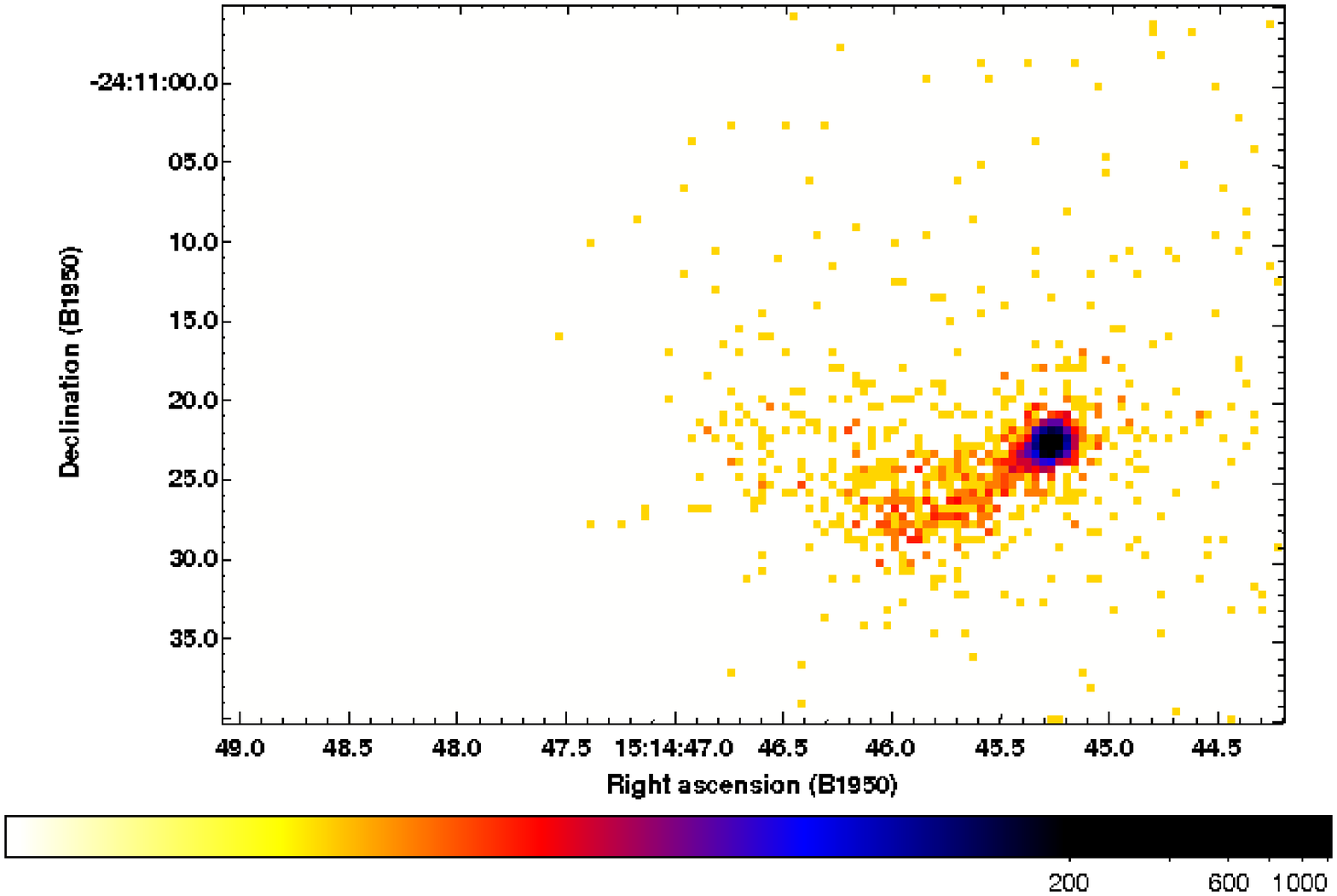}
              \hfil
              \includegraphics[width=0.48\textwidth]{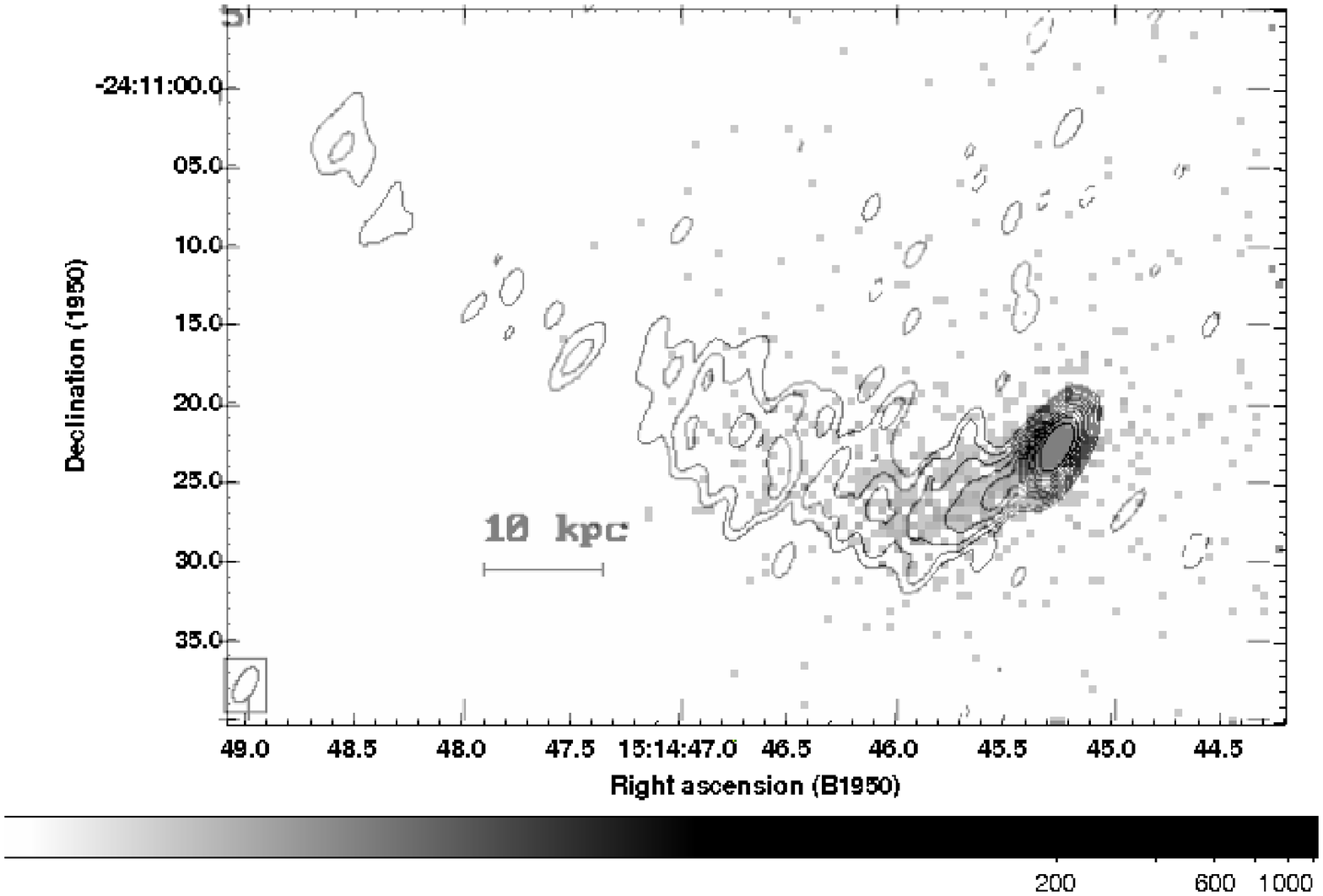}
             }
   \caption{
{\it left}: X-ray count map from 0.2 to 8keV extracted from the 14ks observation by Chandra on AP Lib from 4. July 2003. The non-thermal jet is clearly visible. Due to the timed exposure mode, the observed frame is cutted at the left part of the image. Therefore the real extension of the jet is not measurable.
{\it right}: Overlapped are the radio contours of the VLA observation of AP Lib in A+B configuration at 1.36 GHz. The restoring beam is 3.0x2.0 arcsec in PA $28^\circ$ . The peak flux density is 1625 mJy/beam and the r.m.s. noise on the image is 0.15 mJy/beam \cite{Cassaro1999}.
}
   \label{Xjet}
 \end{figure*}

\subsection{Extended X-ray jet}
AP Lib is the first LBL with detected TeV $\gamma$-ray emission and a clearly visible extended non-thermal X-ray jet. 
We detected the X-ray jet in our own analysis of the 14ks Chandra observation of 4. July 2003 (see Fig.~1). The jet is located in the south-east direction of the source. The jet is dominated by non-thermal emission.
Unfortunately only a small frame of the CCD was used due to the used subarray in the timed exposure mode and the exposure is rather low, so that the real extension of the jet cannot be measured.

The spectrum of the core can be well ($\chi^2/dof=165/147$) described by a power law with $\Gamma = 1.58 \pm 0.04$ taking into account the Galactic absorption of $N_H = 8.36\times 10^{20} \; \rm{cm^{-2}}$ (LAB survey,\cite{Kalberla2005}).
The spectrum of the jet can be described by $\Gamma = 1.76 \pm 0.14$ ($\chi^2/dof = 9/19$), but the statistics is low due to low exposure. The resulting fluxes are $F_{\rm{core,2-10keV}} = (2.9 \pm 0.1) \times 10^{-12} \; \rm{erg \; cm^{-2} \; s^{-1}}$ and $F_{\rm{jet,2-10keV}} = (2.6 \pm 0.3) \times 10^{-13} \; \rm{erg \; cm^{-2} \; s^{-1}}$.
No significant spectral change between the jet and the core spectrum could be determined, but the hardness ratio indicates a slight change between the core and the jet. This change is not continuous along the jet indicating structure in the jet, but the statistics is limited so that no clear statement can be made. 

Swift/XRT observations from 2007 to 2011 cannot resolve the jet and the core+jet spectrum can be described by a photon index of $\Gamma \sim 1.6-1.7$.

The spectra are comparable with the original definition of AP Lib being a low-energy peaked BL Lac object with IC dominance in the X-ray spectrum. The hard X-ray instrument BAT onboard Swift detected a flux of $F(14-150keV)=(1.4\pm 0.3)\times 10^{-11}\;\rm{erg\;cm^{-2}\;s^{-1}}$ within 39 months of observations $[10]$. 


\subsection{Radio jet}
Observations with the Very Large Array (VLA) on AP Lib show clear detection of the radio jet (see Fig.~\ref{Xjet}). The radio jet at 1.36GHz emerge along the SE direction and bend towards NE after a dozen of arcseconds, for a total extend of $\sim 55''$. The observations with the D array at 1.4 GHz show a diffuse emission on arcmin scale on the same side of the jet \cite{Cassaro1999}. 
The comparison of the kpc jet in radio and X-rays reveil the same location of the emission along the SE direction. The bended jet $>$10 kpc was not significantly detected by the Chandra observation due to the used subarray and the short exposure. 
Under the assumption of the SSC models, that the same electron population is producing the synchrotron and inverse Compton emission and since the radio and X-ray spectra of the LBL AP Lib cover the lower energy part of both emission peaks in the SED (see Fig.~3), we expect to see the same jet morphology of the emission region in the radio and X-ray regime.
In VLBA observations, \cite{Pushkarev2009} an opening angle of $7.8^\circ$ was determined. 

\subsection{TeV $\gamma$-ray emitter}
The first VHE $\gamma$-ray emission was detected from AP Lib by the H.E.S.S. Cherenkov telescope array in June/July 2010 \cite{Hofmann2010},\cite{Fortin2011}. It was detected with a significance of $6\sigma$ above 300GeV, which corresponds to $\sim 2\%$ of the flux of the Crab nebula. No significant flux variations were observed during the $\sim 8$ hours of observation.
The TeV spectrum can be described by a power law with $\Gamma = 2.5 \pm 0.2$ \cite{Fortin2011}. 
Due to this VHE detection, AP Lib is the third LBL after BL Lac and S5 0716+716 detected at VHE $\gamma$-rays. Although S5 0716+716 seems more likely an IBL from its SED \cite{Tavecchio2010}.
For most LBLs, the expected emission (following the standard SSC model) in the VHE band is well below the sensitivity of current Cherenkov telescopes and therefore it is amazing to detect them.

\subsection{GeV $\gamma$-ray emission}
A GeV source, 1FGL J1517.8-2423 was detected by the Fermi Gamma-ray space telecope and can be associated with AP Lib. The GeV spectrum can be fit by a power law with a photon index of $\Gamma = 2.1 \pm 0.1$ and result in a flux of $F(100\rm{MeV}-100\rm{GeV})=(5.5\pm 0.6)e-12\;\rm{erg\;cm^{-2}\;s^{-1}}$  \cite{Abdo2010}. No significant flux variation was detected over 11 months of observations.

 \begin{figure*}[!t]
   \centerline{\includegraphics[width=0.48\textwidth]{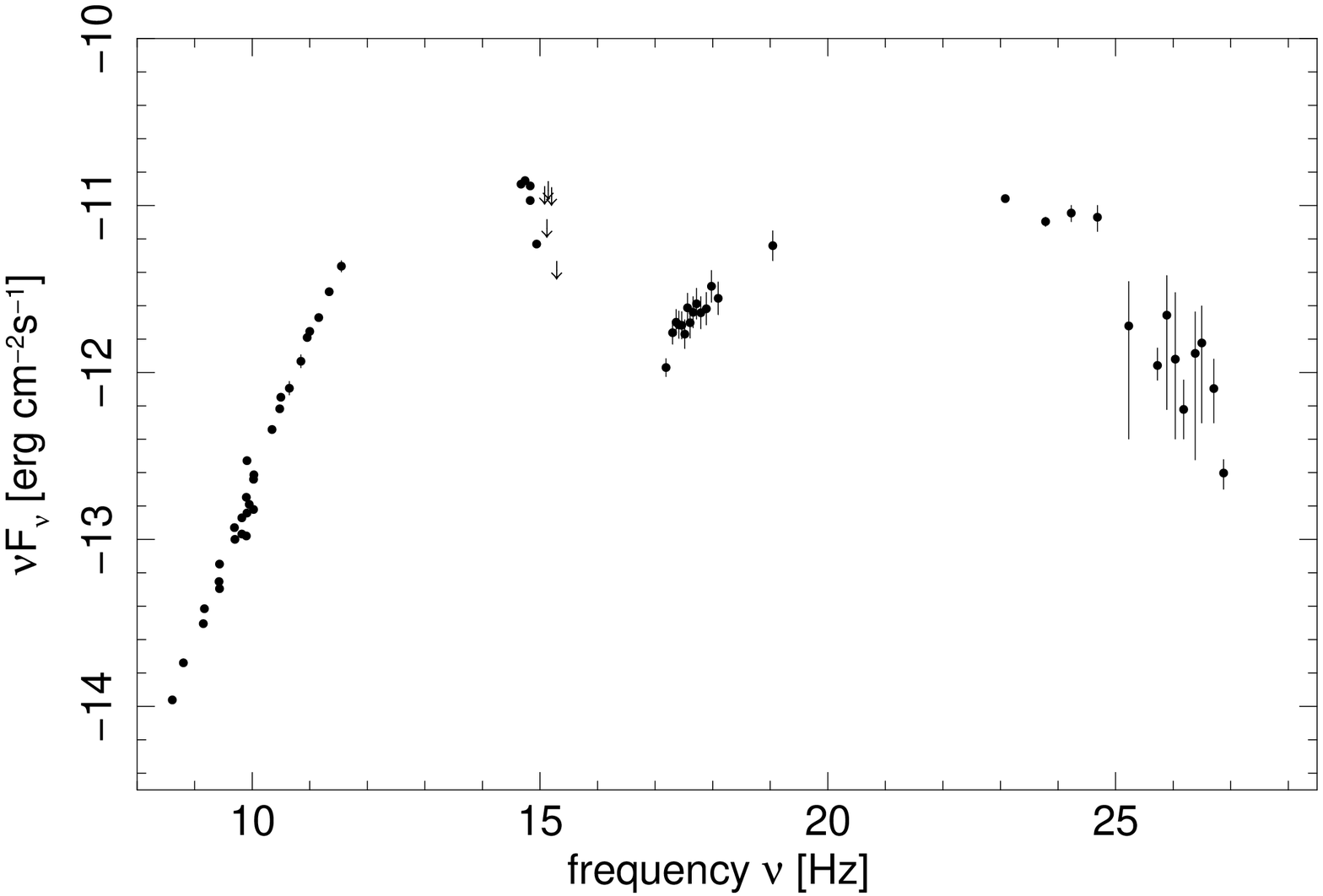}
              \hfil
              \includegraphics[width=0.48\textwidth]{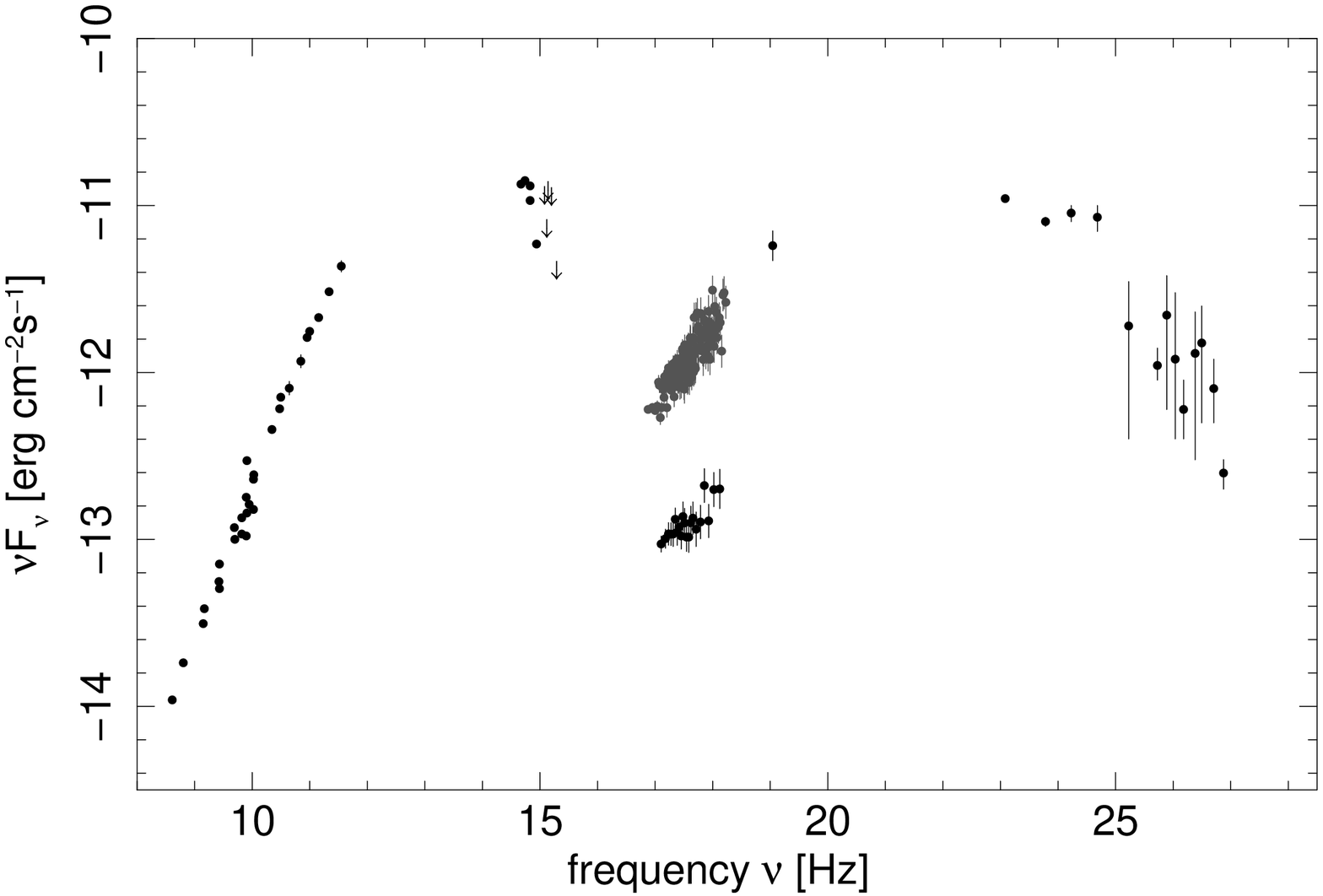}
             }
   \caption{{\it left}: Spectral energy distribution of AP Lib showing historical radio data from \cite{Kuehr}, radio data from PLANCK, optical data from ATOM, UV and X-ray results from the Swift observation in 2011, the hard X-ray flux mentioned in the Swift/BAT catalog \cite{Cusumano2010} and the GeV and TeV flux points 
from \cite{Fortin2011}.
A narrow synchrotron and a very broad inverse Compton peak is visible and the X-ray domain is clearly dominated by the IC emission. Therefore AP Lib is classified as LBL.
{\it right}: In addition the core (grey) and jet (black) spectrum measured by Chandra in 2003 are shown. Both spectra are dominated by the inverse Compton emission of Ap Lib. 
            }
   \label{SED}
 \end{figure*}

\subsection{Optical intra-day variability and morphology}
AP Librae is known to be one of the most active blazars in the optical band. In data from 1989, intra-day varibility was detected with a very high rate of change of $0.06\pm 0.01$ mag/hr \cite{Carini1991}. Even on shorter timescales of 20min, variation of 0.5mag have been detected in 1973 \cite{Miller1974}.

From UBV measurements, \cite{Westerlund1982} a nucleus with 'miniquasar' properties and an extended component that could be an E galaxy with an ultraviolet excess was identified.

In 1993, \cite{Stickel1993} found that the host galaxy of Ap Lib appears asymmetric and elongated towards a nearby galaxy ($\approx 65''$ to the north east). Therefore they suggested that Ap Lib is an interacting system. Further observations reveal that both galaxies are at the same redshift indicating a real association \cite{Pesce1994}.

\section{Variability of the optical, UV and X-ray emission}

The X-ray emission of Ap Lib show only marginal variation between 2003 and 2011. In 2003, Chandra observed the source with a core flux of $F_{\rm{2-10keV}} = (2.9 \pm 0.1)\times 10^{-12} \;\rm{erg \; cm^{-2} \; s^{-1}}$. Swift observations have been conducted in 2007, 2008, 2010 and 2011 and an increase of flux by $\sim 40\%$ was detected between 2008 and 2010 while no spectral change appeared between 2007 and 2011. 

The simultaneous observed UV and optical emission by UVOT onboard Swift satellite, show in the UV bands a slight increase over the time range 2007-2011 by $\sim 0.3 \;\rm{mag}$. The B and V filter observations show constant fluxes. The U band observations show slight increase by $\sim 0.2 \;\rm{mag}$ with similar trend to the UV observations, but a significant drop by $\sim 0.4 \;\rm{mag}$ appeared in July 2010 which unfortunately was not covered by the other UVOT filters. 

\section{Spectral energy distribution}



In Fig.~\ref{SED} a collection of data from radio to TeV $\gamma$-rays are shown to characterize the spectral energy distribution (SED) of Ap Lib. In the radio, observations in several bands are shown from \cite{Kuehr} as well as new results from PLANCK\footnote{e.g. http://www.isdc.unige.ch/heavens\_webapp/integral\\ and http://tools.asdc.asi.it/} are shown. In the optical bands, observations by ATOM are shown, but since the optical emission of the source is very variable, this represents only one flux state of the source. 

The host galaxy was studied with deep NOT observations by \cite{Pursimo2002} resulting in $m_{\rm{host}} = 14.29 \pm 0.01 \;\rm{mag}$ and an effective radius of $r_e = 6.72 \pm 0.04''$ in the R-band. 

In the UV, optical and X-ray regime, the simultaneous observations by Swift are shown. The presented flux values are corrected by the influence of the host galaxy using the template for elliptical galaxies at $z=0$ by \cite{Fukugita1995}. Since the influence by the host galaxy in the UV is unknown, the fluxes are shown as upper limits for the SED characterizing the blazar. The UV and optical fluxes are also corrected fro dust absorption using $E(B-V) = 0.138$ (\cite{Schlegel1998}). The X-ray spectra have been corrected for the Galactic absorption. In near and far UV, results from the GALEX satellite, corrected for extinction, are shown. They are marked as upper limits as well, since they are not taken simultaneous to the UV observations by Swift and the influence of the host galaxy is unknown. 

Together with the Swift spectrum, also the core and jet spectra taken from the Chandra observations are shown here. Clearly a different flux level than with Swift is detected wherefore any interpretation in combination with the shown UV and optical fluxes is not possible. A slight change in spectral shape is visible between the core and the jet spectrum, but both spectra are dominated by inverse Compton emission.

The hard X-ray instrument BAT onboard Swift detected a flux of $F(14-150keV)=(1.4\pm 0.3)\times 10^{-11}\;\rm{erg\;cm^{-2}\;s^{-1}}$ within 39 months of observations $[10]$. 
The GeV flux points originate from the Fermi catalog \cite{Abdo2010}. A narrow synchrotron and a very broad inverse Compton peak is visible and the X-ray domain is clearly dominated by the IC emission. Therefore AP Lib is classified as LBL.

\clearpage

\end{document}